# Intrinsic and Extrinsic Contributions to the Lattice Parameter of GaMnAs

L. X. Zhao, C. R. Staddon, K. Y. Wang, K. W. Edmonds, R. P. Campion, B. L Gallagher, C. T. Foxon

*School of Physics and Astronomy, University of Nottingham, Nottingham NG7 2RD, United Kingdom*

We report on measurements of the crystal structure and hole density in a series of as-grown and annealed GaMnAs samples. The measured hole densities are used to obtain the fraction of incorporated Mn atoms occupying interstitial and substitutional sites. This allows us to make a direct comparison of the measured lattice parameters with recent density functional theory (DFT) predictions. We find that the decrease in lattice constant observed on annealing is smaller than that predicted due to the out diffusion of interstitial Mn during annealing. The measured lattice parameters after annealing are still significantly larger than that of GaAs even in samples with very low compensation. This indicates that the intrinsic lattice parameter of GaMnAs is significantly larger than that of GaAs, in contradiction to the DFT prediction.

The discovery of carrier mediated ferromagnetism in GaMnAs[1] has led to extensive investigations of the properties of these materials and potential spintronic devices based upon them[2, 3, 4]. Recently the realisation that defects in these materials can play a very important role has led to a considerably improved understanding of their electrical and magnetic properties. It has been shown that a substantial fraction of the Mn in as grown GaMnAs can occupy interstitial rather than substitutional sites[5]. It has also been demonstrated that, in thin GaMnAs epi-layers, low temperature annealing can remove most of the interstitial Mn by diffusion to the free surface[6]. $Mn_I$ is a double donor in $Ga_{1-x}Mn_xAs$ and therefore compensates holes provided by substitutional $Mn_{sub}$. Furthermore tight-binding[7] and density-functional[6] calculations indicate that $Mn_I$ couples antiferromagnetically to neighboring $Mn_{sub}$. Interstitial Mn is therefore particularly efficient at suppressing ferromagnetism in GaMnAs and low temperature annealing can lead to large increases in Curie temperature. After annealing compensation can be very low[8] and it has recently been demonstrated that the previously observed "magnetisation deficit" is can also removed by annealing.[9]

It is generally found that the lattice constant of GaMnAs increases with increasing Mn concentration[10]. It had been assumed that this was because GaMnAs has an intrinsic lattice constant which is larger than that of GaAs, and that the measured lattice constant could be used to estimate the Mn content of GaMnAs samples[10]. However it has now been demonstrated that the measured lattice constant, for a given nominal Mn content, can be strongly influenced by the growth condition[11]. Recently Masek et al[12] using a density functional calculation have predicted that the increase in lattice constant due to substitutional Mn should be almost zero and have suggested that the observed increases in lattice constant with increasing Mn concentration is probably all due to defects. They predicted that the lattice constant will be

$$a = a_0 + 0.02x + 0.69y + 1.05z \ (\text{Å}) \quad (1)$$

where $a_0$ is the GaAs lattice parameter and, x, y and z are the fractions of $Mn_{sub}$, $As_{Ga}$ antisites and $Mn_I$. Recently decreases of the lattice parameter of thin GaMnAs layers after annealing have been reported[13,14] and discussed qualitatively in terms of the influence of defects.

Here we present experimental measurements of the hole densities, and crystal structure of a series of as-grown and annealed $Ga_{1-x}Mn_xAs$ thin films along with GaAs control samples. Full details of the growth procedures used are presented elsewhere[15]. A 100nm thick high temperature GaAs buffer layer is first grown on the (100) GaAs substrate at 580°C. A 50nm thick low temperature GaAs layer is then grown followed by a 50nm thick $Ga_{1-x}Mn_xAs$ layer grown at the same low temperature. The temperature used for these layers, which is chosen to ensure two dimensional growth, was 245, 200, and 175°C for the $x$= 0.022, 0.056 & 0.090 samples respectively. Control films were also grown at these same low temperatures with a 100nm thick low temperature GaAs layer and no $Ga_{1-x}Mn_xAs$ top layer. The Mn concentration $x$ was obtained from the Mn / Ga flux ratio calibrated by secondary ion mass spectrometry (SIMS) of 1μm thick films grown under the same conditions. Samples used for electrical measurements were annealed at 190°C until the *in-situ* monitored electrical resistance appeared to reach a minimum[9]. This occurred after about 100 hours at 190°C. Samples used in the structural measurements were annealed for 100 hours. Carrier densities were obtained from Hall measurements at low temperatures (0.3 - 7K) and high magnetic fields (to 16.5T), using a fitting procedure to remove the contribution from the anomalous Hall Effect[9]. The structural properties were determined using a Philips X'Pert Materials Research diffractometer with a primary mirror, a four-bounce Ge (220) monochromator at the incident beam and a secondary triple bounce Ge (220) analyser in front of the detector. Results show that all the films are fully strained with respect to the GaAs substrate. The reciprocal space maps obtained using the 004 symmetric reflections were projected onto the ω-2θ axis, by integration in the ω direction, to give pseudo-rocking curves. To obtain accurate relaxed lattice parameters the ω-2θ curves obtained are compared with simulations using Philips X'pert Epitaxy.

Table 1 gives the measured hole densities for the samples before and after annealing. From these measured values we obtain values for the concentrations of substitutional and interstitial Mn assuming that the compensation of the holes is entirely due to double-donor interstitial Mn. Since the hole density is very close to the substitutional Mn density after annealing for the $x=0.022$ and $0.056$ Mn samples this is a reasonable assumption. For the $x=0.090$ sample significant compensation is still present after annealing. This could be due to either in complete out diffusion of interstitial Mn or As-antisites $As_{Ga}$, which are also double donors. However we can be confident that the observed *change* in compensation on annealing is due to the out diffusion of interstitial Mn because $As_{Ga}$ is stable at much higher temperatures than those employed here. For the $x=0.022$ sample before and after annealing and the $x=0.056$ sample after annealing the hole density is within the estimated uncertainty equal to the $Mn_{sub}$ concentration showing that the number of compensating defects is small in these samples.

Figure 1 shows the experimental and simulated (004) ω/2θ scans for the as grown and annealed 100nm low temperature GaAs control films. For the GaAs films grown at 240°C there is only one single narrow peak and we cannot detect any difference between the lattice constants of the LT GaAs layer and the substrate. For the films grown at lower temperature there is a shoulder at the lower angle side of the main peak and interference fringes showing that the LT GaAs layers have larger lattice constants than the substrate and a different refractive index. The simulation agrees well with the measurements and yields increases in lattice constant of 0.0011Å for 200°C and 0.0018Å for 175°C. The increase in lattice constant in LT GaAs is known to be due to the formation of $As_{Ga}$ antisites[16]. Equation (1) gives $As_{Ga}$ densities of $3.5 \times 10^{19}$ and $6 \times 10^{19}$ cm$^{-3}$ for the LT GaAs layers grown at 200°C and 175°C. These values are very small compared with the Mn and hole densities in the GaMnAs samples. We see no significant differences in the control samples after low temperature annealing. This is consistent with the annealing not affecting the antisite densities.

Figure 2 shows the experimental and simulated ω-2θ scans for the samples with 50nm GaMnAs layers. Comparison of the results for $x$=0.090 Mn with the appropriate control sample shows that the features due to the substrate and LT GaAs layers are largely unchanged and an additional broad peak, due to the thin GaMnAs layer having a significantly larger lattice constant than the GaAs, is present. On annealing a large shift in the positions of the GaMnAs peak is apparent for the 0.090 and 0.056 samples. This shift is to be expected due to the out diffusion of $Mn_I$. In a series of 1μm samples we have studied such shifts are not apparent. This is consistent with the $Mn_I$ not having time to diffuse to the free surface in such thick samples[6]. The simulated results, for which the LT-GaAs lattice constant is kept the same as that of the LT-GaAs control sample, agree very well with the experimental results.

Figure 3 shows that the lattice constant has been reduced substantially by annealing at high Mn concentration but that the lattice constants of the GaMnAs layers after annealing are still much larger than that of GaAs. From electrical measurement, we know that there is almost no compensation in the $x$=0.022 and 0.056 materials after annealing. The difference in lattice constant cannot therefore be due to either $Mn_I$ or $As_{Ga}$. This strongly indicate that, in contradiction to the calculations of Masek et al, the intrinsic lattice constant of GaMnAs is significantly larger than that of GaAs. In figure 3 the best linear fit line through the 0, 0.022, and 0.056 data points after annealing is shown. This gives a dependence of $a=a_o+(0.26\pm0.08)x$. The data points for the annealed 0.090 samples also falls on this line. This is surprising as the compensation is still significant in this sample after annealing. We note that very recently similar large dilation of lattice constant at low compensation has been observed.[17]

The inset of figure 3 shows the measured changes in lattice constant of the GaMnAs due to annealing plotted against the change in interstitial density obtained using the data of table 1. The lattice parameter varies linearly with Mn interstitial density but with a coefficient of 0.6±0.2 rather than the value of 1.05 of equation (1).

In summary, combined measurements of the hole concentration and lattice parameter before and after annealing in a series of (Ga,Mn)As samples has allowed us to make a direct comparison with recent density functional theory (DFT) predictions. We find that annealing at low temperatures has no influence on the density of As antisite defects in low temperature GaAs. Our results show that the lattice constant of GaMnAs varies approximately linearly with interstitial Mn density but with a coefficient which is significantly smaller than the predicted value. We also find that the intrinsic lattice parameter of GaMnAs is significantly larger than that of GaAs, in contradiction to the DFT prediction.

## Acknowledgements

The work was supported by the UK EPSRC and the EC FENIKS project. KWE is supported by the Royal Society (UK). We acknowledge valuable discussions with Tomas Jungwirth.

**Captions:**

Fig. 1 Comparison rocking curves before and after annealing for the samples with 100nm of low temperature GaAs grown at the temperatures indicated. Blue curves are experimental data and red ones are simulation.

Fig. 2 Comparison rocking curves before and after annealing for the samples with 50nm of GaMnAs, with the Mn content indicated, and 50nm of low temperature GaAs. Blue curves are experimental data and red ones are simulation

Fig.3 Relaxed lattice constants for the 50nm thick GaMnAs films before annealing (squares) and after annealing (triangular) as a function of substitutional Mn content. The inset shows the change of the relaxed lattice constant as a function of the change in the fraction of interstitial Mn, (x of equation 1), due to the out-diffusion interstitial Mn during annealing (i.e. $Mn_{I,i} - Mn_{I,f}$).

Table 1. Measured initial hole density, $p_i$, final hole density after annealing, $p_f$, and calculated substitutional Mn concentration, $Mn_{sub}$, and interstitial Mn concentration before, $Mn_{I,i}$ and after, $Mn_{I,f}$ annealing, all in units of $10^{20} cm^{-3}$.

| %$Mn_{total}$ | $p_i$ | $p_f$ | $Mn_{sub}$ | $Mn_{I,i}$ | $Mn_{I,f}$ | $p_i/Mn_{sub}$ | $p_f/Mn_{sub}$ |
|---|---|---|---|---|---|---|---|
| 2.2 | 3.5±0.4 | 4.7±0.4 | 4.4±0.4 | 0.5±0.2 | 0±0.2 | 0.8±0.04 | 1.06±0.09 |
| 5.6 | 4.4±0.4 | 9.8±0.5 | 9.7±0.4 | 2.6±0.2 | 0±0.2 | 0.45±0.04 | 1.01±0.05 |
| 9.0 | 2.5±0.5 | 9.0±1 | 14.0±0.8 | 5.7±0.4 | 2.5±0.4 | 0.18±0.05 | 0.64±0.08 |

Table 1 L. X. Zhao et al Applied Physics Letter

**Figures:**

Fig.1 L. X. Zhao et al Applied Physics Letter

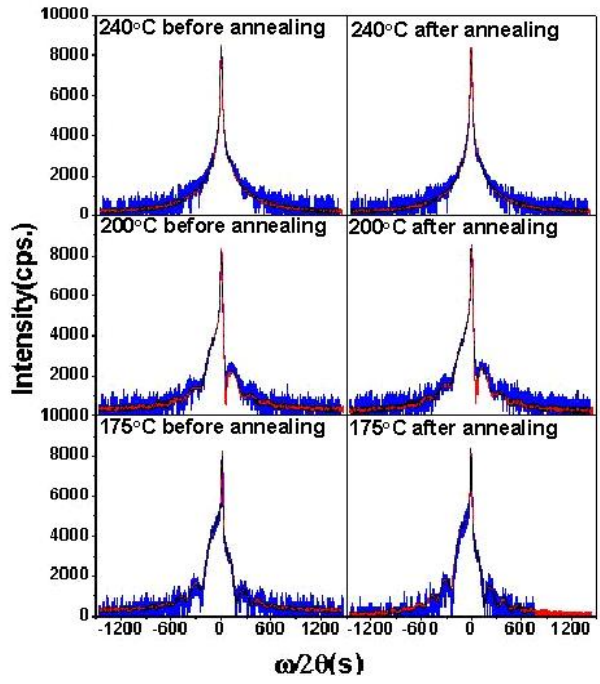

Fig.2 L. X. Zhao et al Applied Physics Letter

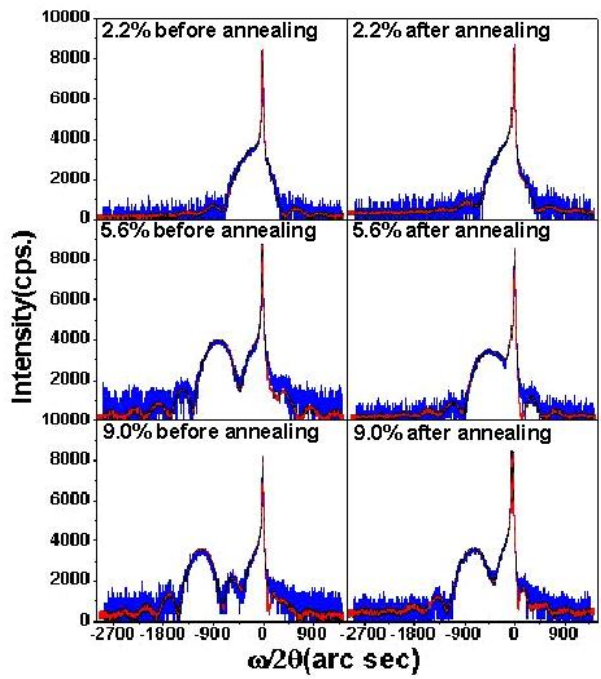

Fig.3 L.X. Zhao et al Applied Physics letter

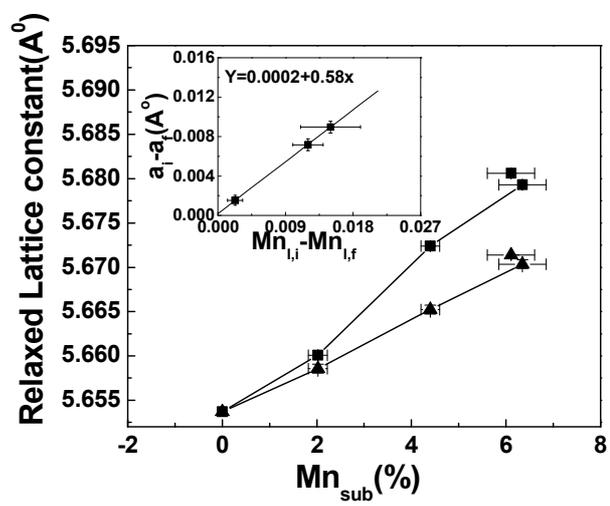